# I-SiamIDS: An Improved Siam-IDS for handling class imbalance in Network-based Intrusion Detection Systems

Punam Bedi, Neha Gupta, Vinita Jindal

## Abstract

Network-based Intrusion Detection Systems (NIDSs) identify malicious activities by analyzing network traffic. NIDSs are trained with the samples of benign and intrusive network traffic. Training samples belong to either majority or minority classes depending upon the number of available instances. Majority classes consist of abundant samples for the normal traffic as well as for recurrent intrusions. Whereas, minority classes include fewer samples for unknown events or infrequent intrusions. NIDSs trained on such imbalanced data tend to give biased predictions against minority attack classes, causing undetected or misclassified intrusions. Past research works handled this class imbalance problem using data-level approaches that either increase minority class samples or decrease majority class samples in the training data set. Although these data-level balancing approaches indirectly improve the performance of NIDSs, they do not address the underlying issue in NIDSs i.e. they are unable to identify attacks having limited training data only. This paper proposes an algorithm-level approach called Improved Siam-IDS (I-SiamIDS), which is a two-layer ensemble for handling class imbalance problem. I-SiamIDS identifies both majority and minority classes at the algorithm-level without using any data-level balancing techniques. The first layer of I-SiamIDS uses an ensemble of binary eXtreme Gradient Boosting (b-XGBoost), Siamese Neural Network (Siamese-NN) and Deep Neural Network (DNN) for hierarchical filtration of input samples to identify attacks. These attacks are then sent to the second layer of I-SiamIDS for classification into different attack classes using multi-class eXtreme Gradient Boosting classifier (m-XGBoost). As compared to its counterparts, I-SiamIDS showed significant improvement in terms of Accuracy, Recall, Precision, F1-score and values of Area Under the Curve (AUC) for both NSL-KDD and CIDDS-001 datasets. To further strengthen the results, computational cost analysis was also performed to study the acceptability of the proposed I-SiamIDS.

## 1. Introduction

In the present digital era, computer networks have become a crucial part of human life. They not only serve as mediums for exchange of digital information, but also act as providers of several different services to their users. This dependence of individuals and organizations on computer networks has made them a lucrative target of cyber-attacks. Cyber criminals try to compromise the confidentiality, integrity and availability of online data and services through different network intrusions. To identify such intrusions, Intrusion Detection Systems (IDSs) came into existence. IDSs monitor and analyse online traffic to segregate normal and malicious content. When IDSs are deployed within a network to identify network-based intrusions, they are known as Network-based Intrusion Detection Systems (NIDSs). These systems capture online network traffic and analyse it to detect the presence of attacks. A significant advantage of using NIDSs is their ability to single-handedly monitor the traffic traversing several devices present within the network.

Different types of NIDSs use different techniques for detecting malicious network traffic. Signature-based NIDSs (S-NIDSs) maintain a record of known attack patterns called signatures. They compare the network traffic with the stored signatures and raise an alarm whenever a match occurs between the two. Though these systems raise very few false alarms, they are unable to identify new types of network intrusions whose signatures are not contained in the attack repository. Unlike S-NIDSs, Anomaly-based NIDSs (A-NIDSs) develop a profile of normal network traffic and then compare the online network traffic to the developed normal profile. This allows them to identify known, as well as novel attacks by flagging any deviation from the normal profile as an intrusion (Deka, Kalita,

Bhattacharya, & Kalita, 2015). Since A-NIDSs are effective against both known and unknown attacks, we design and develop an A-NIDS in this paper.

Before deploying an A-NIDS in real-world, it must be trained on sample data that adheres to the characteristics of real-world network traffic. Majority of this network traffic is benign in nature, apart from rare events that generate malicious content. Intrusion detection datasets that are developed using real networks or emulated network environments, also have a similar distribution of benign and malicious network traffic. Such datasets, that have a significant difference in the number of samples of different classes, are known as imbalanced datasets. In an imbalanced dataset, the class(es) having a majority share of total samples is (are) called the majority class(es), while the class(es) having a minority share of total samples is (are) called the minority class(es). In imbalanced intrusion detection datasets, majority classes consist of benign samples and frequent attack samples, while minority classes consist of infrequent attack samples. As given by the authors in their paper (Ring, Wunderlich, Scheuring, Landes, & Hotho, 2019), many such imbalanced datasets are available for developing A-NIDSs. This paper utilises NSL-KDD and CIDDS-001 intrusion detection datasets for experimentation purposes.

In NSL-KDD dataset, the normal samples, Denial of Service (DoS) attack samples and Probe attack samples represent the majority classes, while Remote-to-Local (R2L) and User-to-Root (U2R) samples form the minority classes of the dataset. Similarly, CIDDS-001 dataset contains normal, DoS attacks and Port Scan attacks as the majority classes, while Ping Scan and Brute Force attacks form the minority classes of the dataset. A-NIDSs tend to be biased against minority classes of the imbalanced intrusion detection dataset. This is because most of these A-NIDSs utilize Machine Learning (ML) algorithms which require a large number of training samples to learn the characteristics of different classes of data (Çavuşoğlu, 2019). The lack of training samples for minority classes leads to an increase in the number of unidentified intrusions. These are referred to as False Negatives. To overcome this problem, two possible approaches exist: data-level approaches and algorithm-level approaches. Data-level techniques try to reduce the imbalance ratio in the dataset by either increasing or decreasing the number of samples present in different classes. On the other hand, algorithm-level approaches try to develop algorithms that can effectively handle the imbalance among classes without the need of any external data modification techniques.

In A-NIDSs, most researchers have utilised one of the three data-level techniques namely, Oversampling, Undersampling and Synthetic Minority Oversampling Technique (SMOTE). Oversampling increases the number of minority class samples by duplicating them, undersampling reduces the number of majority class samples by eliminating them and SMOTE creates synthetic samples of the minority class. Though these techniques balance the dataset, each of them has some disadvantage. Oversampling leads to overfitting of the minority class samples, Undersampling causes loss of important information from the majority class samples and the synthetic samples generated by SMOTE may not be true representatives of the minority class. Hence, there is a need to devise algorithms that can be used to develop A-NIDSs in such a way, that they are able to correctly classify minority attack classes. The authors in their work have reviewed different algorithms for performing multi-class classification on imbalanced data (Bi & Zhang, 2018). They also proposed Diversified Error Correcting Output Codes for tackling the data imbalance problem. The authors compared the performance of their proposed method with the existing techniques using publicly available class-imbalanced datasets.

In (Bedi, Gupta, & Jindal, 2019), the authors utilised Siamese Neural Network (Siamese-NN) to handle the class imbalance problem. Their system, named Siam-IDS, was able to classify minority class samples without using any dataset balancing technique. Though Siam-IDS achieved acceptable Recall values, the Precision values obtained by it were low. In this paper, we propose an algorithm-level approach for handling class imbalance, named Improved Siam-IDS (I-SiamIDS). The proposed I-SiamIDS is a two-layer ensemble of eXtreme Gradient Boosting (XGBoost), Siamese-NN and Deep Neural Network (DNN). I-SiamIDS identifies larger number of attacks and handles the class imbalance problem more efficiently than Siam-IDS. The first layer of I-SiamIDS separates the benign and malicious samples using a hierarchical filtration process. The second layer classifies the attacks identified by the first layer into respective attack categories. The proposed I-SiamIDS outperforms Siam-IDS as well as four other multi-class classifiers namely DNN, Convolutional Neural Network (CNN), Random Forest (RF) and XGBoost. An improvement in the Accuracy, Recall, Precision, and F1-score for multi-class classification and Area Under Curve (AUC) value for binary classification, was observed on both NSL-KDD and CIDDS-001 datasets by using the proposed I-SiamIDS. This highlights the effectiveness of I-SiamIDS in identifying intrusions without the use of data balancing techniques. Furthermore, computational cost analysis was also used as a comparison indicator to study the acceptability of the proposed I-SiamIDS in this paper.

The remaining paper is organized as follows: Section 2 describes DNN, Siamese-NN, XGBoost, NSL-KDD dataset and CIDDS-001 dataset. Section 3 presents the literature review. Section 4 gives the details of the proposed I-SiamIDS. Section 5 explains the experimental work and the results obtained. Section 6 concludes the paper.

## 2. Background Information

This section presents a brief description of the algorithms and the datasets that have been used in the development of proposed I-SiamIDS system. These include Deep Neural Network, Siamese Neural Network, eXtreme Gradient Boosting algorithm, NSL-KDD dataset and CIDDS-001 dataset.

*Deep Neural Network*

A Deep Neural Network (DNN) is a ML algorithm whose structure is inspired by the interconnected architecture of neurons found in human brain. Every DNN consists of three types of layers: input layer, output layer and hidden layer(s). Each layer is made up of several neurons and each neuron has a connection with all the neurons present in the succeeding DNN layer, but no connections exist between neurons present in the same layer. Moreover, the inter-layer connections are present only in the forward direction without the presence of any backlinks or loops. This makes DNN a fully-connected feed-forward network. Each feed-forward connection is associated with a weight value. The training phase of the DNN aims to find the best set of weights using the backpropagation algorithm. In each of the training iteration, an input is given to the input layer.

Each neuron present in the input layer multiplies the input with the weight associated with it and forwards it to all the neurons of the first hidden layer. The neurons of the hidden layer apply a non-linear activation function to the weighted input that they receive from the preceding layer and forward it to the next layer. Each hidden layer enables the DNN to learn more significant information about the input. The last hidden layer feeds the processed information to the output layer, which outputs the probability of the input belonging to each output category. If the label with the highest output probability does not match with the actual label, the output is propagated back towards the initial layers and the weights are adjusted. Once the training is complete, the final weights are then used during the testing phase to classify the test sample into one of the classes (Gupta, Bedi, & Jindal, 2019).

*Siamese Neural Network*

Siamese Neural Network (Siamese-NN) was first used for signature verification in (Bromley, Guyon, LeCun, Sickinger, & Shah, 1994). It is a Few-shot learning algorithm that uses the concept of the input similarity for performing classification (Chowdhury, et al., 2017). Siamese-NN accepts a pair of inputs and outputs the similarity score between them. The similarity score is calculated using a distance function between the feature representations of the two inputs. These feature representations are computed by two identical neural networks, which have the same set of weights. The presence of identical networks ensures that the feature representations of the input pair do not change by varying the order of inputs. Siamese-NN has proved its mettle in several research areas such as age estimation

(Jeong, Lee, Park, & Park, 2018), video-based person re-identification (Liu, Sun, Xu, Xu, & Yu, 2019) and human gait prediction (Zhang, Liu, Ma, & Fu, 2016). It was also utilised by (Bedi, Gupta, & Jindal, 2019) to develop an A-NIDS that handled the class imbalance issue in intrusion detection system. In the proposed system, identical DNNs were used to calculate the feature representations of the inputs to the Siamese-NN. Further, Euclidean distance was used as the similarity function to compute the distance between the calculated feature representations.

*eXtreme Gradient Boosting (XGBoost)*

XGBoost is a ML technique developed by Tianqi Chen in (Chen & Guestrin, 2016). It is an ensemble method which combines multiple weak learners to create a strong learner. In XGBoost, a weak learner is a decision tree which tries to reduce the misclassifications made by the previous decision tree. Classification And Regression Trees (CARTs) are mostly used for this purpose. Each iteration in the training phase builds a weak learner to predict the target variable. The difference between the true value and the predicted value is called the residual error of the iteration. The next decision tree takes these residual errors as the target values and makes the predictions. These predicted values are combined with the predictions made by the previous decision tree. This is done to ensure that every subsequent decision tree minimizes the error of the previous decision tree.

XGBoost aims to optimize the objective function which consists of two parts: the loss function and the regularization function. The loss function minimizes the error and the regularization function prevents the overfitting of the model. XGBoost has an advantage of being faster and more accurate than simple Gradient Boosting. In the proposed I-SiamIDS, XGBoost classifier has been used in both the layers. In the first layer, binary XGBoost (b-XGBoost) has been used for segregating benign and malicious samples. In the second layer, multi-class XGBoost (m-XGBoost) has been used for classifying attacks into their respective classes.

*NSL-KDD dataset*

The NSL-KDD (Network Socket Layer – Knowledge Discovery in Databases) dataset was derived from KDD intrusion detection dataset by (Tavallaee, Bagheri, Lu, & Ghorbani, 2009). The authors removed redundant and duplicate records from the KDD dataset to create the NSL-KDD dataset (Gupta, Bedi, & Jindal, 2019). Two separate CSV files were developed for training and testing purposes. There exist 1,25,973 samples in the training file and 22,544 samples in the testing file. Each file consists of forty-one attributes and a class label. Three attributes contain categorical values while the remaining attributes have numerical values. The class labels also have categorical values: one corresponding to normal samples and the rest corresponding to different attack types.

There are 22 different attack types in the training data. All these attack types can be grouped together into four main attack categories as specified in Table 1. The normal class, DoS attack class and the Probe attack class forms the majority classes of the training data set, while the R2L and U2R classes represent the minority classes of the training data set. The percentage of each attack category in both training and testing datasets has been given in Table 1. The uneven distribution of samples in NSL-KDD dataset makes it a suitable choice for training an A-NIDS to identify both attacks: majority attacks as well as minority attacks.

Table 1: Description of NSL-KDD dataset

| | NSL-KDD Dataset | | | |
|---|---|---|---|---|
| | Training Data | | Testing Data | |
| | Samples | % | Samples | % |
| **Normal** | 67343 | 53.45 | 9711 | 43.07 |
| **DoS** | 45927 | 36.45 | 7458 | 33.08 |
| **Probe** | 11656 | 9.25 | 2421 | 10.73 |
| **R2L** | 995 | 0.007 | 2887 | 12.80 |
| **U2R** | 52 | 0.0004 | 67 | 0.002 |

Table 2: Description of CIDDS-001 dataset

| | CIDDS 001 Dataset | | | |
|---|---|---|---|---|
| | Training Data | | Testing Data | |
| | Samples | % | Samples | % |
| **Normal** | 53000 | 53.17 | 15000 | 56.77 |
| **DoS** | 36000 | 36.12 | 6604 | 24.99 |
| **Port Scan** | 9117 | 9.15 | 3250 | 12.30 |
| **Ping Scan** | 500 | 0.50 | 765 | 2.90 |
| **Brute Force** | 1055 | 1.06 | 803 | 3.04 |

*CIDDS-001 dataset*

The CIDDS-001 (Coburg Intrusion Detection Data Sets) dataset is a unidirectional NetFlow dataset developed in 2017 by (Ring, Wunderlich, Grüdl, Landes, & Hotho, 2017). It comprises of benign and malicious network traffic that was generated and captured by emulating a business environment using OpenStack virtual environment together with an External Server connected to the Internet. The dataset comprises of 3,12,87,934 flows captured in OpenStack environment and 6,71,241 flows captured from External Server. Both these files consist of normal samples and 4 types of attacks namely, Denial of Service, Port Scan, Ping Scan and Brute Force.

CIDDS-001 consists of 11 attributes along with 4 types of labelling attributes. Each of the four attack categories contains different sub-attack classes that are uniquely identified by attack identifiers. For the four main attack classes specified before, a total of 70 sub-attack categories exist in the CIDDS-001 data set. Due to the large size of the dataset, a subset of samples was selected for experimental purposes in this paper. The subset reflects the imbalanced nature of the original CIDDS-001 and includes all the 70 sub-attack categories captured in the OpenStack environment. The details of the selected samples are provided in Table 2.

## 3. Literature Review

This section presents an overview of recent research works in the field of network-based intrusion detection. The author of (Rodda, 2018) analyzed the effectiveness of multi-layer perceptron and radial-basis function for designing a NIDS for multi-class classification. Though multi-layer perceptron gave better results than its counterpart, it was unable to handle minority attack classes viz. R2L and U2R. The accuracy for both these attack classes was found to be 0. A Deep Learning (DL) based NIDS was developed by the authors of (Gurung, Ghose, & Subedi, 2019). Their system utilized sparse Auto-Encoder (AE) for feature learning along with logistic classifier for identifying attacks from NSL-KDD dataset. But the authors only tested their approach for binary classification of samples, without any class-wise evaluation of their system.

In (Shenfield, Day, & Ayesh, 2018), binary classification of network traffic was performed by using DNN. Deep packet inspection was utilized by the authors to detect shellcode patterns in the data. The authors of (Mazini, Shirazi, & Mahdavi, 2019) used a hybrid approach to develop an A-NIDS. The Adaboost algorithm was used as a classifier on two datasets namely, NSL-KDD and ISCXIDS 2017. Artificial Bee Colony algorithm was used for extracting the most important features. In (Idhammad, Afdel, & Belouch, 2018), the authors only identified Distributed DOS attacks using entropy estimation, extra-trees algorithm, Information Gain and co-clustering techniques.

A review of class imbalance in different application areas was presented in (Ali, Shamsuddin, & Ralescu, 2015). The authors described the two main strategies for handling the class imbalance problem: the data-level approach and the algorithm-level approach. The first approach tries to balance the ratio of different classes in data by using a pre-processing technique. Common data-level techniques include Oversampling, Undersampling and variations of SMOTE. The second approach fine-tunes the classification algorithms to increase their capability of detecting minority classes. The authors divided this approach into five major categories, namely one-class learning, improved algorithm, cost sensitive learning, ensemble and hybrid technique. The authors also presented suitable evaluation measures to evaluate the effectiveness of classification in domains with class imbalance. Data-level approaches for handling class imbalance problem were also explored in (Tyagi & Mittal, 2020). Various Undersampling and Oversampling methods were briefly discussed. Five publicly available imbalanced datasets were evaluated using k-Nearest Neighbor (kNN), Neural Network (NN) and SVM. The results indicated that Undersampling methods were more efficient in reducing class imbalance as compared to Oversampling techniques. A variation of SMOTE, namely Density Based SMOTE was developed as an Oversampling technique by (Bunkhumpornpat, Sinapiromsaran, & Lursinsap, 2012) to handle class imbalance.

(Zhou, Yang, Fujita, Chen, & Wen, 2020) presented a fault diagnosis method by utilizing a global optimization Generative Adversarial Network (GAN) for imbalanced datasets. The authors designed a new generator using a traditional backpropagation NN and AE to generate fault features. In addition, a hierarchical discriminator was also proposed which incorporated a DNN fault diagnosis model to the traditional discriminator. (Zhang, et al., 2019) proposed a software for performing multi-class classification on imbalanced data. The authors described different algorithms present in the package along with the latest developments in the area of class imbalance. To handle the class imbalance problem in enterprise credit evaluation, (Sun, Lang, Fujita, & Li, 2018) developed a new Decision Tree (DT) ensemble method by combining SMOTE and Bagging techniques with differentiated sampling rates.

A solution to the class imbalance problem for financial distress prediction was presented by (Sun, Li, Fujita, Fu, & Ai, 2019). Adaboost based SVM with Time Weighting was integrated with SMOTE in two different ways to create a balanced dataset for performing correct predictions. (Wang, et al., 2014) explored the risk associated with different permissions in Android applications. They also studied the usefulness of various permissions in detecting malicious Android applications. Similarly, in (Wang, et al., 2019), a detailed description of features used for tracing malicious applications was presented. The authors also categorized existing works depending on the type of features utilized in them for identifying such applications.

The area of network intrusion detection also suffers from class imbalance problem. In (Abdulhammed, Faezipour, Abuzneid, & AbuMallouh, 2018), the authors compared the performance of five algorithms on the original imbalanced CIDDS-001 dataset and the CIDDS-001 dataset after balancing it. Four dataset balancing techniques namely Up-sampling, Down-sampling, Spread sub-sample and Class Balancer were used for this purpose. It was found that out of DNN, RF, Variational AE, Voting and Stacking algorithms, best results were obtained from RF in most of the cases. The authors of (Xiao & Xiao, 2019) used a modified architecture of Residual Networks, named Simplified Residual Networks, to create an IDS using NSL-KDD dataset. They carried out their experiments after balancing the dataset using the Random Oversampling technique. This method randomly duplicates minority class samples but causes overfitting of data.

The work (Chawla, Bowyer, Hall, & Kegelmeyer, 2002) introduced the concept of SMOTE for handling the class imbalance problem. SMOTE balances the minority class(es) by creating new samples that are similar to existing minority class samples. This technique was used in (Hamid, Sugumaran, & Journaux, 2016). After balancing the KDD99 dataset, the authors performed feature selection and feature extraction to identify the most important features. These features were input to SVM for binary classification of attacks and normal samples. No details were provided for the efficiency of this classifier in detecting minority attack types.

The authors of (Kar, Banerjee, Mondal, Mahapatra, & Chattopadhyay, 2019) applied a similar technique for balancing the minority class. Synthetic samples were generated for each minority class using kNN and edited nearest neighbor approaches. Recent works in the field of network intrusion detection have utilized different tree-based ensemble approaches. The work (Tao, Peng, Zhao, Zhao, & Wang, 2018) made use of Isolation Forests (IFs) to develop an A-NIDS using Spark. In (Dhaliwal, Nahid, & Abbas, 2018), XGBoost was used to develop an IDS for binary classification of normal and attack samples. The authors trained and tested their system on the NSL-KDD dataset and evaluated its performance using different evaluation metrics. XGBoost was also utilized as a detection method in software-defined network-based cloud by the authors of (Chen, et al., 2018). The combination of XGBoost and Adaboost algorithms was used to design a NIDS in (Verma, Anwar, Khan, & Mane, 2018). The efficiency of this pair was tested with and without the use of clustering algorithm for the segregation of benign and malicious network traffic. In (Kaja, Shaout, & Ma, 2019), the authors compared the performance of Naïve Bayes, J48 DT, RF and Adaboost algorithms to classify the attacks identified through k-Means clustering. Though RF gave the best overall accuracy out of the four classifiers, attack-wise evaluation was missing from the results.

In (Lee & Park, 2019), an unsupervised method of reducing the imbalance among the classes of CICIDS 2017 dataset was proposed. It utilized GANs to generate samples similar to the existing minority classes. The performance of RF classifier was measured before and after balancing the dataset. GANs were also used in (Lee, Lim, & Noh, 2020) to generate traffic similar to the minority classes of the datasets. Three intrusion detection datasets were used for this purpose. These included NSL-KDD, ISCX 2012 and USTC_TFC 2016 datasets. The authors noted an increase of 10-12 percent in overall classification accuracy for CNN. Apart from GANs, researchers have also tried using other DL algorithms to reduce the imbalance between majority and minority classes of the datasets. The authors of (Wan, Zhang, & He, 2017) proposed the use of Variational Auto-Encoders to generate synthetic data samples. This technique was tested using the MNIST handwritten digit dataset and affNIST dataset which is developed by applying affine transformation on MNIST dataset.

The class imbalance problem is a major research area in network intrusion detection domain. But most of the research in this area utilizes data-level approaches such as data Oversampling and Undersampling to tackle this issue. Although data-level solutions pave way for proper training of NIDS in an offline environment, they do not make NIDSs capable of detecting new and rare events in real-time scenario with few available training samples. The need of the hour is to develop efficient algorithms that can filter novel as well as infrequent network traffic from huge volumes of benign traffic and known attack patterns. Since limited data is available in such cases, hence the research community must focus on developing algorithm-level approaches to handle imbalanced network traffic without modifying the number of available training samples.

Another drawback of past research works is that they use Accuracy of the classifier as the evaluation criterion to measure its effectiveness. Accuracy is not a true representative of the efficiency of the classifier in case of imbalanced datasets. This is because most of the samples present in imbalanced datasets belong to the majority class. Since classifiers are good at predicting classes with abundant training samples, majority predictions (belonging to the majority class) are correct. This results in high Accuracy even when the classifier incorrectly predicts most of the minority class samples.

In (Bedi, Gupta, & Jindal, 2019), Siamese-NN was used by the authors for handling class imbalance in NIDSs. Siamese-NN uses distance-based approach to identify minority class samples without the use of any data-level balancing technique. The performance of this technique was evaluated using Recall and Precision values. Though Siamese-NN based IDS, named Siam-IDS, achieved high Recall values as compared to DNN and CNN, their Precision values were low. This paper improves the work of (Bedi, Gupta, & Jindal, 2019) and proposes I-SiamIDS, a two-layer algorithm-level approach to effectively identify both majority and minority classes. Accuracy, Recall, Precision and F1 values are used to evaluate the effectiveness of I-SiamIDS over NSL-KDD and CIDDS-001 datasets. Receiver Operating Characteristics (ROC) curve has also been plotted and the corresponding AUC values have been computed for binary classification. In addition to this, computational cost in terms of execution time has been calculated for the proposed system.

I-SiamIDS differs from Siam-IDS in two major aspects as discussed below. The architecture of Siam-IDS consists of a single Few-Shot Learning algorithm (Siamese-NN) for intrusion detection. On the other hand, I-SiamIDS uses an ensemble of Few-Shot Learning, DL and Boosting algorithm (Siamese-NN, DNN and XGBoost respectively) for improved intrusion detection. Using these classifiers, I-SiamIDS performs attack identification at the first layer and attack classification at the second layer. A dedicated layer for attack identification minimizes the number of unidentified intrusions through hierarchical filtration of input samples. The attacks identified at the first layer are classified into different attack classes by the second layer of the I-SiamIDS.

In contrast to I-SiamIDS, Siam-IDS accomplishes both these tasks in a single layer through multi-class classification performed by Siamese-NN. A major drawback of Siam-IDS is that the samples classified as normal by Siamese-NN do not undergo further assessment by any other classifier to identify False Negative predictions. In addition, during experimentation it was found that multi-class XGBoost used in I-SiamIDS, is better at performing attack classification as compared to Siamese-NN used in Siam-IDS. Due to its two-layer ensemble architecture and choice of classifiers, I-SiamIDS proves to be more efficient in detecting intrusions belonging to both majority and minority classes. The proposed I-SiamIDS is described in the next section.

## 4. Proposed I-SiamIDS

In this paper, we propose I-SiamIDS, which is a two-layer ensemble NIDS for identification and classification of intrusions. The first layer of the proposed system performs binary classification for the separation of malicious traffic from benign traffic. It aims to reduce the misclassification of malicious traffic by filtering the benign traffic multiple times so that minimal number of attacks goes undetected. To select an appropriate set of algorithms for this purpose, seven different ML algorithms were tested, which could be combined with Siamese-NN. These algorithms included k-Means clustering, k-NN, RF, IF, b-XGBoost, DNN and CNN. To train and test these algorithms for performing binary classification, the pre-processed datasets namely, NSL-KDD and CIDDS-001, were utilized with binary labels.

After training, each of them was tested using the pre-processed testing datasets having binary test labels. Figure 1 shows the performance Accuracy of all the algorithms on NSL-KDD and CIDDS-001 datasets respectively. The selection of an algorithm was based on its performance Accuracy, which is computed using True Positive (TP), False Positive (FP), True Negative (TN) and False Negative (FN) values generated by the algorithm. Each of these terms have been described below and represented as confusion matrix in Figure 2.

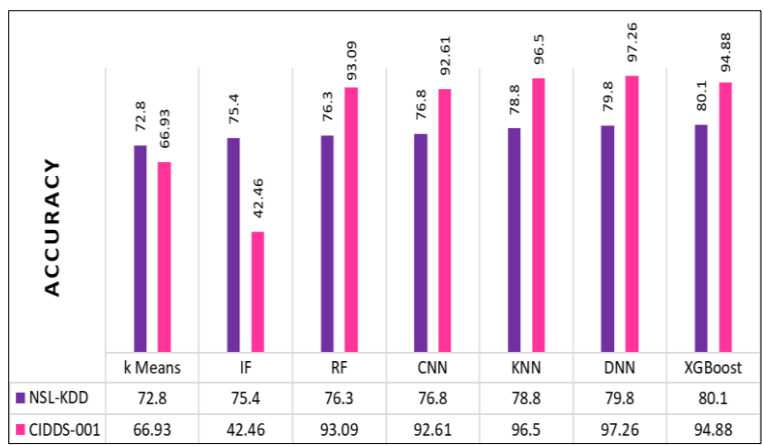
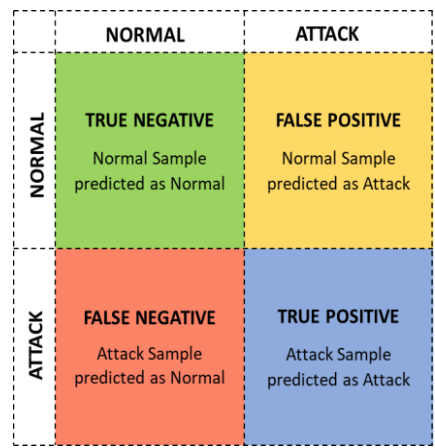

Figure 1: Accuracy of classifiers on NSL-KDD and CIDDS-001 datasets             Figure 2: Confusion Matrix

True Negative: A normal/benign traffic sample which is *correctly* categorized as benign by the IDS.
True Positive: An attack/malicious traffic sample which is *correctly* categorized as attack by the IDS.
False Negative: An attack sample which is *incorrectly* categorized as normal by the IDS.
False Positive: A benign traffic sample which is *incorrectly* categorized as attack by the IDS.

The formula for calculating Accuracy is given by equation (1).

$$Accuracy = \frac{TP+TN}{TN+TP+FP+FN} \qquad (1)$$

Although an efficient IDS must aim to maximize TPs and TNs along with minimizing FPs and FNs, reducing the number of FNs is more crucial than FPs. This is because FPs indicate benign traffic that was classified as malicious by the IDS and leads to false alarms being raised by the IDS. On the other hand, FNs refer to the malicious traffic that escaped the eyes of the IDS and no alarms were raised for it. Since unidentified intrusions are much more dangerous than false alarms, therefore reducing the number of FNs is more important.

Out of the seven aforementioned algorithms, we selected two algorithms having the maximum accuracy. For NSL-KDD dataset, b-XGBoost and DNN were the top performers. In case of CIDDS-001 data, DNN and KNN performed the best. However, KNN being a lazy learner becomes time consuming for large datasets and it becomes unsuitable for intrusion detection in real world scenario. Therefore, instead of selecting KNN, the third best performer i.e. b-XGBoost was selected for Layer 1 of I-SiamIDS. Hence, an ensemble of Siamese-NN, b-XGBoost and DNN was finalised for the first layer of the proposed I-SiamIDS. The reason for selecting more than one classifier was to increase the number of correctly identified attacks and reduce the number of unidentified attacks through hierarchical filtration of benign traffic using multiple classifiers.

Experiments were conducted by considering different number of classifiers for Layer 1. Starting from a single classifier, the number of classifiers was increased till an improvement in the result was seen. It was observed that when more than three classifiers were used, only marginal improvement was achieved. Since, adding more classifiers also leads to an increase in overhead; this cost was much higher than the benefit received from adding the new classifier. Therefore, we found the choice of selecting three classifiers at Layer 1 to be optimal. Moreover, different classifiers were chosen because selecting the same classifier with the best performance more than once, may not lead to good results. This is because a specific classifier may be biased against a specific class of samples. Due to this, the same classifier would never correctly identify that specific category of test data. This was confirmed in the intermediate results that were obtained in the process of model creation. A combination of different classifiers, as selected in this paper, creates a model that combines the strengths of different classifiers and gives improved results.

After the selection of the three classifiers, the next step involved finding the best permutation of these classifiers for Layer 1 of the proposed I-SiamIDS system. In each permutation, the first classifier takes as input a test sample from the pre-processed testing dataset, and classifies it as normal (0) or an attack (1). If the test sample is classified as normal by the first classifier, it is then passed to the second classifier present in the permutation. If the second classifier categorises the input test sample as belonging to the normal class, then this test sample is forwarded to the third classifier. If the last classifier predicts the test sample as normal, then it is accepted to be normal. Otherwise, if any of the three classifiers classify the input test sample as an attack, it is directly sent to Layer 2 of the proposed system. Each of the six permutations P1, P2, P3, P4, P5 and P6 formed by the three classifiers, tests each sample of the pre-processed testing dataset in this manner.

For NSL-KDD and CIDDS-001 datasets, the results obtained by each permutation have been shown in Figure 13 and Figure 14 in the Appendix section. It was observed that though the intermediate results were different, the combined result obtained by each permutation was exactly the same for each corresponding dataset. Hence, it was inferred that the order of the classifiers is independent of the result obtained at the first layer. Since all the permutations of Layer 1 classifiers were giving same results, therefore we selected $P_5$ (XGBoost => Siamese-NN => DNN) as the final permutation for the classifiers at Layer 1. Figure 3 shows the diagrammatic representation of the proposed I-SiamIDS.

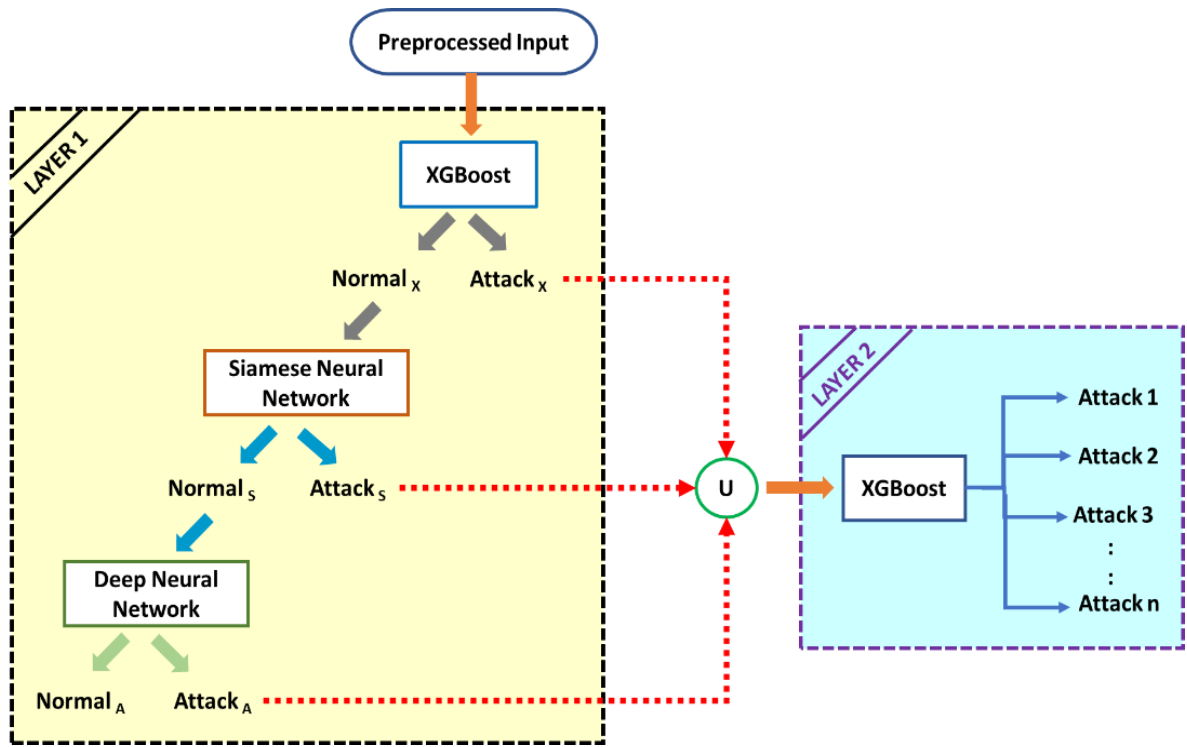

Figure 3: Proposed I-SiamIDS

The second layer of I-SiamIDS was designed to classify the attacks identified at Layer 1 into specific categories. For NSL-KDD dataset, these attack categories include DoS, Probe, R2L and U2R, whereas for CIDDS-001 dataset, the attack categories are DoS, Port Scan, Ping Scan and Brute Force. This type of segregation into different attack types is important for the network administrator who can take appropriate responsive steps depending on the type of the intrusion detected. To select an appropriate classifier for this purpose, two classifiers namely m-XGBoost and Siamese-NN were trained using NSL-KDD and CIDDS-001 training datasets containing multi-class labels. After training the two classifiers on the datasets, their performance was evaluated using the testing datasets. It was found that m-XGBoost performed better than Siamese-NN and therefore it was selected as the Layer 2 classifier. The algorithm for the proposed I-SiamIDS is shown in Figure 4.

```
Algorithm for proposed I-SiamIDS

Input:   Testing dataset T_s
Output:  Predicted labels for each test sample present in T_s

//      Layer 1
1.      Let N_X and A_X be two empty sets
2.      For each test sample t_i in T_s
3.          predict t_i using b-XGBoost and assign the predicted label to p_i
4.          if p_i is normal
5.              add sample t_i to set N_X
6.          else
7.              add sample t_i to set A_X
8.      end

9.      Let N_S and A_S be two empty sets
10.     For each test sample t_j in N_X
11.         predict t_j using Siamese-NN and assign the predicted label to p_j
12.         if p_j is normal
13.             add sample t_j to set N_S
14.         else
15.             add sample t_j to set A_S
16.     end

17.     Let N_A and A_A be two empty sets
18.     For each test sample t_k in N_S
19.         predict t_k using ANN and assign the predicted label to p_k
20.         if prediction p_k is normal
21.             add sample t_k to set N_A
                // p_k is the final predicted label for test sample t_k
22.         else
23.             add sample t_k to set A_A
24.     end

//      Layer 2
25.     Let S be the union of A_X, A_S, A_A sets
26.     For each test sample t_s in S
27.         predict t_s using m-XGBoost and assign the predicted label to p_s
            // p_s represents the label corresponding to one of the attack classes
            // p_s is the final predicted label for test sample t_s
28.     end
```

Figure 4: Algorithm for proposed I-SiamIDS

## 5. Experiments and Results

The proposed two-layer ensemble system was developed using an Intel® Core™ i7-8750H processor with Windows 10 operating system. Python programming language was used for implementing the proposed I-SiamIDS. Two intrusion detection datasets namely NSL-KDD and CIDDS-001 were used for experimentation. The development process of I-SiamIDS began with dataset pre-processing. The NSL-KDD training and testing dataset contains forty-one features, out of which three features have categorical values while others have numerical values. On the other hand, there are five categorical features in the CIDDS-001 dataset. Also, in both the datasets, the values in various numerical features span different numeric ranges. This type of data, having a mix of feature types and value ranges, cannot be input to ML algorithms directly. Both the datasets must be processed prior to their use in the training and testing stages respectively. To remove this asymmetry, dataset pre-processing was performed in two steps: Quantization and Normalization. Each of these steps has been described in the following sub-sections.

*Quantization*

The NSL-KDD dataset contains three categorical features, namely *protocol, service* and *flag*. Since ML algorithms cannot process categorical feature values, the quantization step converted these three categorical features into numerical features. This was done by assigning a unique natural number corresponding to each category present in the categorical feature. This process was repeated for each of the three categorical features present in training and testing datasets. Once all the three categorical features were converted to numeric features, the dataset consisted of features with numerical values only. The same steps were followed to quantize the CIDDS-001 dataset which contains five categorical attributes namely, *Date first seen*, *Proto*, *Src IP Addr*, *Dst IP Addr* and *Flags*. Moreover, the 42$^{nd}$ attribute of NSL-KDD dataset and 11$^{th}$ attribute of CIDDS-001 dataset also contain categorical class labels. These labels were converted to binary values (for Layer 1 processing) by labelling *normal* label as 0 and all other attack labels as 1.

*Normalization*

In the second step of dataset pre-processing, the values of all the features of the quantized NSL-KDD and CIDDS-001 datasets were normalized to bring them in a uniform range of [0, 1]. If $v_{old}$ refers to the un-normalized value of a feature $f_i$, having $v_{max}$ as the maximum value and $v_{min}$ as the minimum value, then the corresponding normalized value $v_{new}$ is given by the formula in equation (2).

$$v_{new} = \frac{v_{old} - v_{min}}{v_{max} - v_{min}} \qquad (2)$$

After the completion of the two steps of dataset pre-processing, the pre-processed datasets were used for developing the proposed two-layer ensemble system. The first layer of I-SiamIDS is an ensemble of three classifiers namely b-XGBoost, Siamese-NN and DNN (used in this specific order). Siamese-NN consists of two DNNs for computing the feature representations of the input pair. Each DNN comprises of an input layer, four hidden layers of 1024, 512, 256, 128 neurons respectively and an output layer of 64 neurons. In addition, a dropout layer with a dropout factor of 0.5 is used before every hidden layer and the output layer. The distance between the feature vectors computed by the two DNNs is calculated using Euclidean distance and the contrastive loss function is used to minimize the error during training via the Adam optimizer. The DNN used at Layer 1 consists of an input layer, five hidden layers having 1024, 512, 256, 128, 64 neurons respectively and an output layer of 2 neurons for binary classification. Hyperbolic Tangent is used as the activation function in the hidden layers. A dropout layer with a dropout factor of 0.1 is used between each of the hidden layers and the output layer.

While testing I-SiamIDS, each network traffic sample first passes through b-XGBoost classifier, which classifies it as either normal or attack. The sample that is classified as normal by b-XGBoost is then passed through the second classifier i.e. Siamese-NN. This classifier analyzes the incoming sample and performs binary classification to segregate normal and attack traffic. This allows the identification of malicious traffic that was misclassified as benign by the previous XGBoost classifier. To further prevent any malicious traffic from escaping the IDS, the traffic that is classified as normal by Siamese-NN is passed to DNN for binary classification. If DNN also classifies its input as benign, then it is considered to be normal without any further evaluation. The traffic that is reported as malicious by any of the three classifiers at Layer 1 is passed to Layer 2 of the proposed IDS. Layer 2 classifies the malicious traffic into one of the attack categories. This multi-class classification is performed by m-XGBoost classifier which is trained to categorize the input into multiple attack classes. This categorization of attacks into their respective classes allows for exact recognition and response for each attack category.

The proposed two-layer ensemble system was tested using the NSL-KDD dataset and CIDDS-001 dataset. In NSL-KDD, each of the 22,544 test samples was input to b-XGBoost classifier present at Layer 1. Out of all these input samples, b-XGBoost classified 13,513 samples as benign/normal and 9031 as malicious. All the normal samples were sent to the second classifier of Layer 1 i.e. Siamese-NN. While 10,722 samples out of 13,513 input samples were classified as normal by Siamese-NN, it predicted 2791 samples as being anomalous. The samples classified as benign by the second classifier were forwarded to the last classifier of the permutation i.e. DNN. Out of the 10,722 samples that were input to it, DNN categorized 10,530 samples as benign and 192 samples as malicious.

Similarly, for CIDDS-001 dataset, 26,422 test samples were input to b-XGBoost at Layer 1. Out of these, 11,812 were predicted as malicious and the remaining 14,610 were sent to Siamese-NN for classification. 13,562 of these instances were classified as normal and the rest 1048 were flagged malicious. From a total of 13,562 test samples received by DNN, 13,203 were considered to be benign, while 359 were predicted as attack samples. The test samples that were categorized as attacks by the ensemble of classifiers present at Layer 1, were sent to Layer 2 for the segregation of attacks into their respective categories. The m-XGBoost at Layer 2 classifies the attacks into DoS, Probe, R2L and U2R attack categories for NSL-KDD dataset and Port Scan, DoS, Ping Scan and Brute Force attack categories for CIDDS-001 dataset. To evaluate the performance of the proposed system with respect to its counterparts DNN, CNN, XGBoost, RF and Siam-IDS, three evaluation metrics namely Recall, Precision and F1-score have been calculated for multi-class classification.

Recall is the ratio of the number of samples of a class that were correctly identified, to the total number of samples belonging to that class. If class A contains $n_A$ number of samples and out of these $n_A$ samples, if only $n_p$ samples were correctly identified, then the formula for Recall can be written as in equation (3).

$$Recall = \frac{n_p}{n_A} \quad (3)$$

Precision is the ratio of the number of samples that actually belong to a class, to the total number of samples that were predicted as belonging to that class. If $n_{pA}$ represents the number of samples that were predicted as class A samples and out of these $n_{pA}$ samples, only $n_{aA}$ samples actually belong to class A, then the formula for Precision can be written as in equation (4).

$$Precision = \frac{n_{aA}}{n_{pA}} \quad (4)$$

F1-score refers to the harmonic mean of Recall and Precision values. It is an evaluation metric that gives equal weightage to both Recall and Precision scores. Its formula is given by equation (5).

$$F1\text{-}score = \frac{2}{\frac{1}{Recall} + \frac{1}{Precision}} \quad (5)$$

The Recall, Precision and F1-scores obtained by all the algorithms on NSL-KDD dataset, have been shown in Figure 5, Figure 6 and Figure 7 respectively. It can be seen from Figure 5 that I-SiamIDS achieved higher Recall values for all the attack classes as compared to its counterparts. This result clearly indicates that I-SiamIDS is capable of identifying a greater number of intrusions, for both majority and minority classes, as compared to other standard classifiers. Similarly, the Precision values of the proposed I-SiamIDS were close behind the Precision values obtained by its five competitors. Moreover, the Precision obtained by I-SiamIDS for normal class is the highest among all other classifiers. In case of F1-values, I-SiamIDS's results were the best among all classifiers for Normal, Probe, R2L and U2R classes of the dataset. For DoS attack, its value was close behind the F1-values for other classifiers.

It must be noted that Precision reflects the percentage of correct predictions made for a specific class by a classifier. So, even if the denominator is small i.e. the total number of predictions for a specific class is very less, then also the Precision score will be high. On the other hand, even if both the numerator and denominator are high i.e. a large number of predictions are correct, then also the Precision score can be low. This is the reason behind high Precision values of DNN and CNN for minority classes and comparatively low Precision values of I-SiamIDS for U2R and R2L classes.

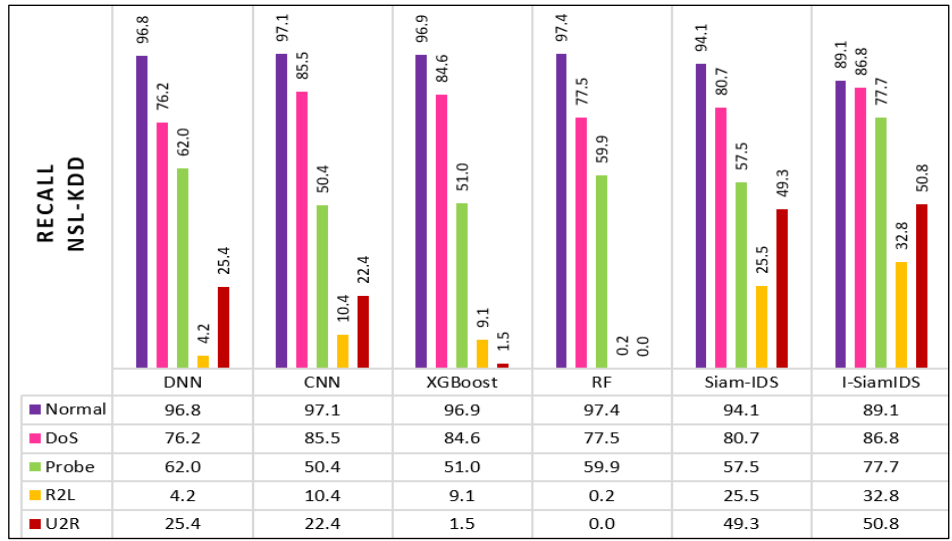

Figure 5: Recall values obtained on NSL-KDD dataset

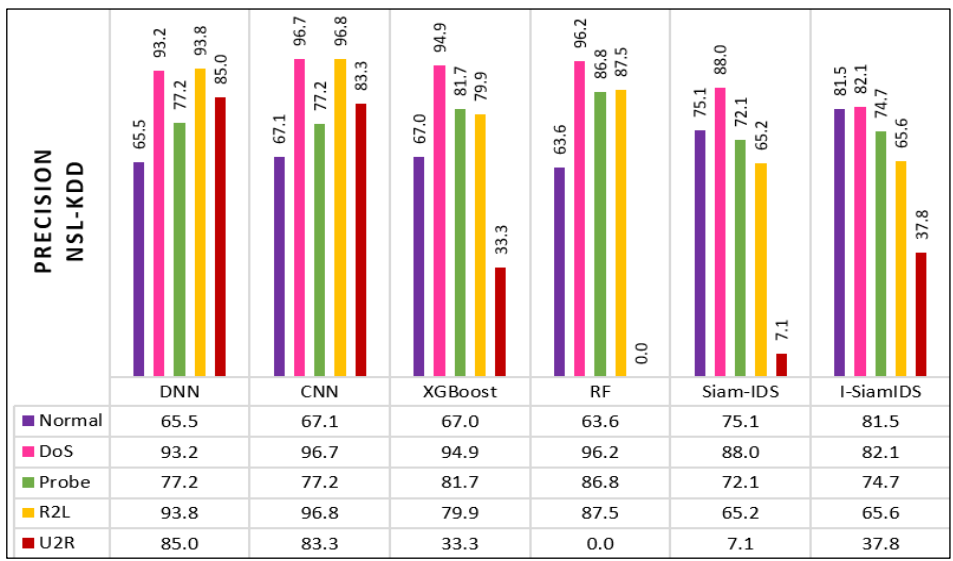

Figure 6: Precision values obtained on NSL-KDD dataset

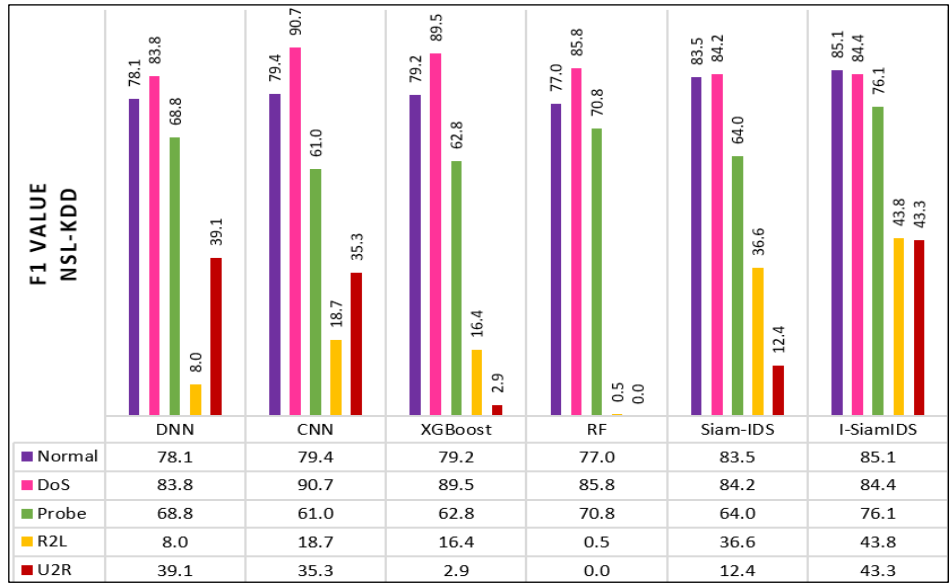

Figure 7: F1-values obtained on NSL-KDD dataset

The Recall, Precision and F1-scores obtained by all the algorithms on CIDDS-001 dataset have been shown in Figure 8, Figure 9 and Figure 10 respectively. I-SiamIDS achieved the second highest Recall value for PingScan and Brute Force classes among other algorithms. The Recall values of I-SiamIDS for the remaining three categories of this dataset were also among the top three values for each class. In addition, the proposed technique achieved the highest Precision values for Normal, Port Scan and Brute Force categories. In the remaining two categories, its Precision scores were the second best as compared to its competitors. Furthermore, the F1-scores obtained by our proposed technique are the best for Normal and Brute Force classes. The results achieved by I-SiamIDS for DoS, PortScan and Ping Scan attacks were among the top three among all the competitors.

Since the F1-values obtained by I-SiamIDS on both the datasets are higher than all or most of the other benchmark ML and DL algorithms, it is safe to conclude that the proposed I-SiamIDS is efficient as a NIDS for both majority and minority attack classes. The values obtained for I-SiamIDS for different evaluation metrics, are highly promising on both the datasets. This highlights our proposed method's efficiency in handling the class imbalance issue for network-based intrusion detection.

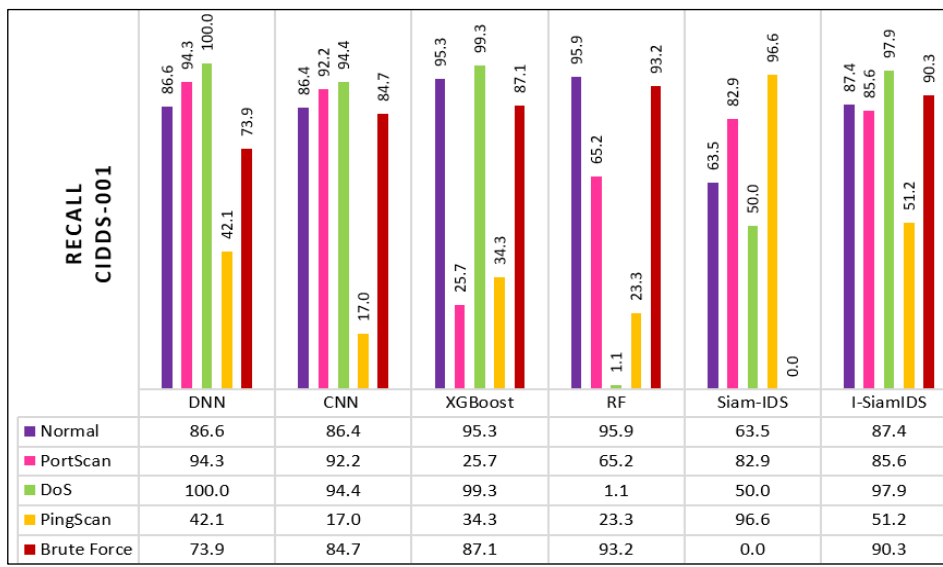

Figure 8: Recall values obtained on CIDDS-001 dataset

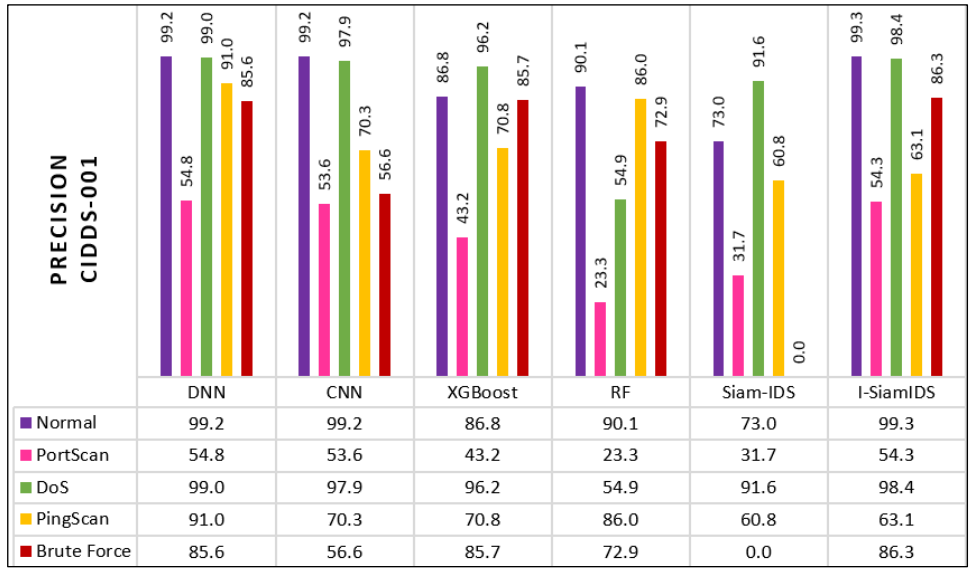

Figure 9: Precision values obtained on CIDDS-001 dataset

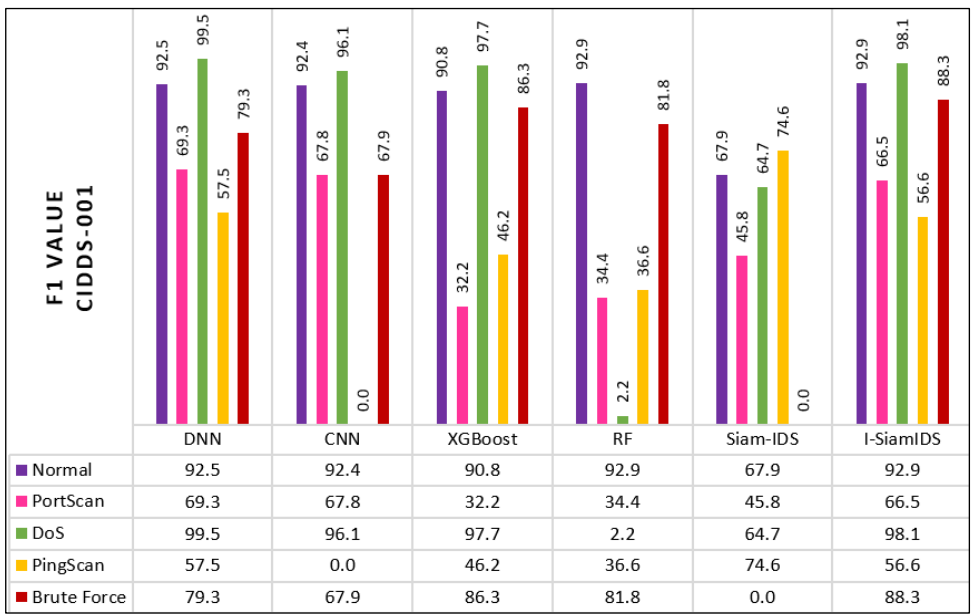

Figure 10: F1-values obtained on CIDDS-001 dataset

In addition to the above-mentioned metrics, ROC curves have also been plotted for evaluating I-SiamIDS's ability to perform binary classification. ROC curve is drawn using False Positive Rate on the horizontal axis and True Positive Rate on the vertical axis for different threshold values. The area under the ROC curve, known as AUC value, reflects the effectiveness of the classifier in performing binary classification. Higher value of AUC indicates that the model is highly efficient in distinguishing the output classes.

In this paper, ROC curves have been plotted by performing binary classification on normal and attack class of both the datasets. Figures 11(a)-(d) depict the ROC curves for I-SiamIDS and its five counterparts. Figure 11(a) and Figure 11(b) present the ROC curve for normal and attack samples of NSL-KDD dataset respectively. Similarly, Figure 11(c) and Figure 11(d) present the ROC curve for normal and attack samples of CIDDS-001 dataset respectively. The AUC values corresponding to all the ROC curves have been shown in Table 3.

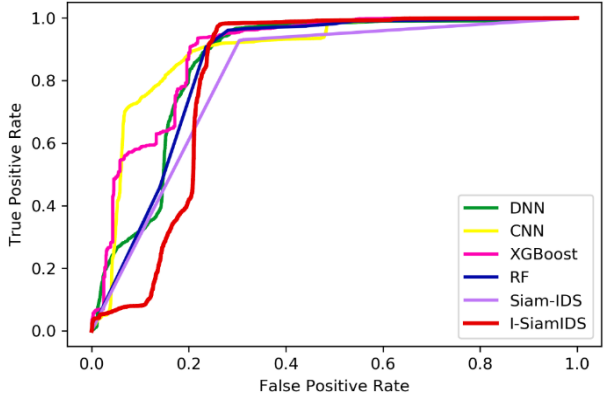
Figure 11(a): ROC curve for normal samples of NSL-KDD dataset

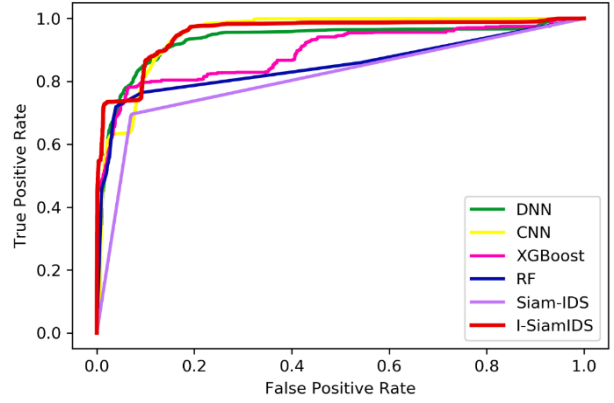
Figure 11(b): ROC curve for attack samples of NSL-KDD dataset

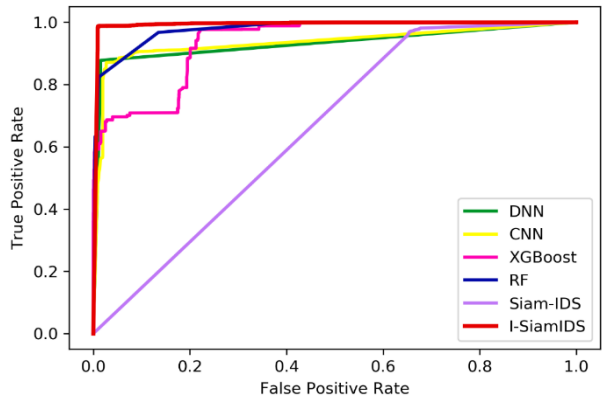
Figure 11 (c): ROC curve for normal samples of CIDDS-001 dataset

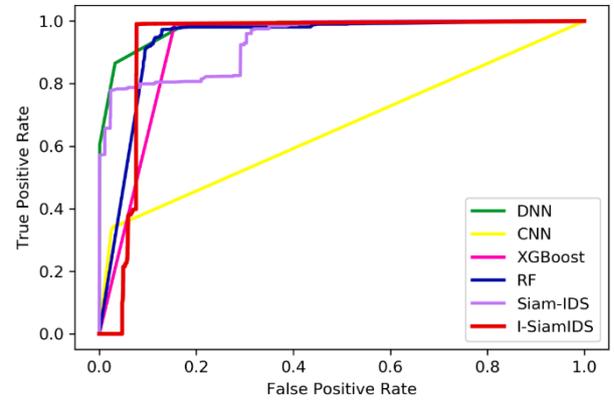
Figure 11 (d): ROC curve for attack samples of CIDDS-001 dataset

Table 3: AUC values for binary classification on NSL-KDD and CIDDS-001 datasets

| Dataset / Classifier | NSL-KDD Normal | NSL-KDD Attack | CIDDS-001 Normal | CIDDS-001 Attack |
|---|---|---|---|---|
| **DNN** | 0.85 | 0.93 | 0.93 | 0.91 |
| **CNN** | 0.89 | 0.94 | 0.93 | 0.93 |
| **XGBoost** | 0.89 | 0.89 | 0.93 | 0.93 |
| **RF** | 0.84 | 0.84 | 0.97 | 0.97 |
| **Siam-IDS** | 0.80 | 0.81 | 0.65 | 0.65 |
| **Proposed I-SiamIDS** | 0.81 | 0.95 | 0.99 | 0.93 |

As shown in Table 3, the proposed I-SiamIDS achieves the highest AUC value when identifying attack samples present in NSL-KDD dataset. This highlights its effectiveness in correctly identifying intrusions present in online network traffic. The performance of I-SiamIDS with respect to normal samples of the NSL-KDD dataset is also satisfactory and it is not very far behind its counterparts. The reason behind this slightly lower value is the multi-level filtration of normal samples performed by I-SiamIDS in the first layer. Hierarchical filtration of normal samples through multiple classifiers ensures that no attack sample escapes undetected. For CIDDS-001 dataset, I-SiamIDS outperforms all other classifiers by achieving the highest AUC value for normal samples. In case of attack samples present in the CIDDS-001 dataset, the proposed I-SiamIDS becomes the second-best candidate after RF, in terms of its AUC score. However, it must be noted that I-SiamIDS outperforms RF in terms of Recall, Precision and F1-values for multi-class classification on CIDDS-001 dataset. The aforementioned results clearly depict that I-SiamIDS is highly efficient in segregating normal and malicious network traffic.

Another important parameter that is crucial for a NIDS is its computational cost. There are two ways in which the computational cost of a model can be calculated: either in terms of floating-point operations, or in terms of execution time. However, the first approach fails to consider various other operational costs that cannot be directly mapped to the total number of floating-point operations (Justus, Brennan, Bonner, & McGough, 2018). Due to this reason, it is rarely adopted by researchers as a measure of computing cost. On the contrary, several authors have utilised execution time to calculate the computational costs of their model (Bonfitto, Feraco, Tonoli, Amati, & Monti, 2019), (Justus, Brennan, Bonner, & McGough, 2018), (Laudani, Lozito, Fulginei, & Salvini, 2015). Therefore, this paper also uses execution time (a.k.a. testing time) as a comparison indicator for the proposed I-SiamIDS. Though training time can also be used for comparison, but it plays a less significant role as compared to testing time. This is because, even if the training time of a NIDS is high, this time will only be required once before the NIDS is deployed in the real world. So, in this paper, we are calculating testing time for measuring computational cost.

To calculate the average testing time for the normal and attack class, ten random samples from both the classes were selected from NSL-KDD and CIDDS-001 datasets. I-SiamIDS and the other five classifiers were evaluated on each of these samples. For each classifier, the testing time of all the samples was recorded. Then, the average testing time per normal sample and the average testing time per attack sample were computed on both the datasets. For NSL-KDD dataset, it was observed that the average testing time per normal sample was 0.0019 seconds for DNN, 0.0043 seconds for CNN, 1.0379 seconds for XGBoost, 0.1286 seconds for RF, 0.0156 seconds for Siam-IDS and 0.4395 seconds for the proposed I-SiamIDS. For the same dataset, the average testing time per attack sample was 0.0019 seconds for DNN, 0.0033 seconds for CNN, 1.0345 seconds for XGBoost, 0.1197 seconds for RF, 0.0169 seconds for Siam-IDS and 0.9551 seconds for I-SiamIDS. It was observed that the time taken by I-SiamIDS is less than the time taken by XGBoost because, I-SiamIDS uses b-XGBoost in the first layer and m-XGBoost in the second layer. The m-XGBoost used in proposed system only segregates different attack classes as compared to XGBoost (used for comparison) that classifies normal as well as attack classes of the datasets.

Further experiments showed that for CIDDS-001 dataset, the average testing time per normal sample was 0.0019 seconds for DNN, 0.0041 seconds for CNN, 0.9634 seconds for XGBoost, 0.0020 seconds for RF, 0.1064 seconds for Siam-IDS and 0.4345 seconds for the proposed I-SiamIDS. The average testing time per attack sample was 0.0019 seconds for DNN, 0.0029 seconds for CNN, 1.0387 seconds for XGBoost, 0.0019 seconds for RF, 0.1049 seconds for Siam-IDS and 1.0720 seconds for the proposed I-SiamIDS. From the above-mentioned time requirements, it can be seen that the average testing time of I-SiamIDS is slightly higher than its counterparts. But it must be noted that even if the proposed I-SiamIDS requires more testing time, it ensures that its attack detection capability is high and minimal number of attack samples goes undetected. The same is also verified by the high accuracy values achieved for multi-class classification by I-SiamIDS. Figure 12 (a) and Figure 12 (b) depict the accuracy values obtained by I-SiamIDS and its five counterparts on NSL-KDD and CIDDS-001 datasets. Hence, it can be concluded that the proposed I-SiamIDS is capable of identifying intrusions in a time-bound manner, which makes it a strong candidate for an efficient NIDS.

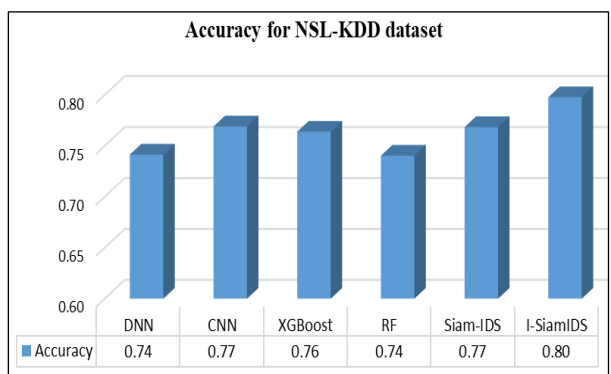
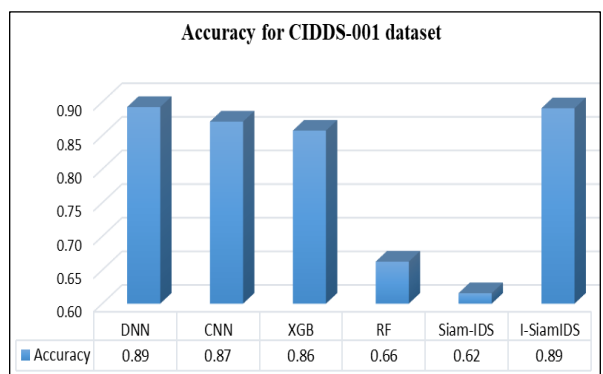

Figure 12 (a): Accuracy on NSL-KDD dataset    Figure 12 (b): Accuracy on CIDDS-001 dataset

## 6. Conclusion

A major challenge in the development of Network-based Intrusion Detection Systems (NIDSs) is the presence of imbalanced network traffic. This traffic consists of a large number of samples for benign and/or recurrent intrusions and limited number of samples for unknown events and/or infrequent intrusions. An efficient NIDSs must be able to identify all types of intrusions by handling this class imbalance in network traffic. In this paper, we proposed Improved Siam-IDS (I-SiamIDS), which is a two-layer ensemble for handling the problem of class imbalance using an algorithm-level approach. I-SiamIDS uses an ensemble of binary eXtreme Gradient Boosting, Siamese Neural Network and Deep Neural Network classifiers at the first layer. This layer performs hierarchical filtration of network data into benign and malicious samples. Filtration of incoming data multiple times through different classifiers minimizes the chances of malicious traffic going undetected by I-SiamIDS. The attack samples identified at the first layer were input to the second layer of I-SiamIDS comprising of multi-class eXtreme Gradient Boosting for classification into four main attack categories. I-SiamIDS was trained and tested using the NSL-KDD and CIDDS-001 datasets without using any data-level techniques of balancing the dataset. Its performance was evaluated against IDSs developed using Deep Neural Network, Convolutional Neural Network, Random Forest, eXtreme Gradient Boosting and Siam-IDS classifiers. It was observed that I-SiamIDS achieved higher Accuracy, Recall, Precision, F1 scores and AUC values as compared to the five algorithms in consideration. Further analysis based on computational cost also showed the acceptabilty of the proposed I-SiamIDS in terms of execution time. All these results highlight the effectiveness of I-SiamIDS in detecting attacks in an imbalanced network environment as compared to its counterparts.

# Appendix

## Permutation 1

| Siamese-NN | Normal | Attack |     | DNN    | Normal | Attack |     | XGBoost | Normal | Attack |     | P1     | Normal | Attack |
|------------|--------|--------|-----|--------|--------|--------|-----|---------|--------|--------|-----|--------|--------|--------|
| Normal     | 8823   | 888    |     | Normal | 8712   | 111    |     | Normal  | 8652   | 60     |     | Normal | 8652   | 1059   |
| Attack     | 2828   | 10005  |     | Attack | 2321   | 507    |     | Attack  | 1878   | 443    |     | Attack | 1878   | 10955  |

## Permutation 2

| Siamese-NN | Normal | Attack |     | XGBoost | Normal | Attack |     | DNN    | Normal | Attack |     | P2     | Normal | Attack |
|------------|--------|--------|-----|---------|--------|--------|-----|--------|--------|--------|-----|--------|--------|--------|
| Normal     | 8823   | 888    |     | Normal  | 8682   | 141    |     | Normal | 8652   | 30     |     | Normal | 8652   | 1059   |
| Attack     | 2828   | 10005  |     | Attack  | 2040   | 788    |     | Attack | 1878   | 162    |     | Attack | 1878   | 10955  |

## Permutation 3

| DNN    | Normal | Attack |     | Siamese-NN | Normal | Attack |     | XGBoost | Normal | Attack |     | P3     | Normal | Attack |
|--------|--------|--------|-----|------------|--------|--------|-----|---------|--------|--------|-----|--------|--------|--------|
| Normal | 9419   | 292    |     | Normal     | 8712   | 707    |     | Normal  | 8652   | 60     |     | Normal | 8652   | 1059   |
| Attack | 4257   | 8576   |     | Attack     | 2321   | 1936   |     | Attack  | 1878   | 443    |     | Attack | 1878   | 10955  |

## Permutation 4

| DNN    | Normal | Attack |     | XGBoost | Normal | Attack |     | Siamese-NN | Normal | Attack |     | P4     | Normal | Attack |
|--------|--------|--------|-----|---------|--------|--------|-----|------------|--------|--------|-----|--------|--------|--------|
| Normal | 9419   | 292    |     | Normal  | 9310   | 109    |     | Normal     | 8652   | 658    |     | Normal | 8652   | 1059   |
| Attack | 4257   | 8576   |     | Attack  | 3332   | 925    |     | Attack     | 1878   | 1454   |     | Attack | 1878   | 10955  |

## Permutation 5

| XGBoost | Normal | Attack |     | Siamese-NN | Normal | Attack |     | DNN    | Normal | Attack |     | P5     | Normal | Attack |
|---------|--------|--------|-----|------------|--------|--------|-----|--------|--------|--------|-----|--------|--------|--------|
| Normal  | 9374   | 337    |     | Normal     | 8682   | 692    |     | Normal | 8652   | 30     |     | Normal | 8652   | 1059   |
| Attack  | 4139   | 8694   |     | Attack     | 2040   | 2099   |     | Attack | 1878   | 162    |     | Attack | 1878   | 10955  |

## Permutation 6

| XGBoost | Normal | Attack |     | DNN    | Normal | Attack |     | Siamese-NN | Normal | Attack |     | P6     | Normal | Attack |
|---------|--------|--------|-----|--------|--------|--------|-----|------------|--------|--------|-----|--------|--------|--------|
| Normal  | 9374   | 337    |     | Normal | 9310   | 64     |     | Normal     | 8652   | 658    |     | Normal | 8652   | 1059   |
| Attack  | 4139   | 8694   |     | Attack | 3332   | 807    |     | Attack     | 1878   | 1454   |     | Attack | 1878   | 10955  |

Figure 13: Permutations of Layer 1 classifiers for NSL-KDD dataset

## Permutation 1

| Siamese-NN | Normal | Attack |   | DNN | Normal | Attack |   | XGBoost | Normal | Attack |   | P1 | Normal | Attack |
|---|---|---|---|---|---|---|---|---|---|---|---|---|---|---|
| Normal | 14097 | 903 |   | Normal | 13588 | 509 |   | Normal | 13107 | 481 | = | Normal | 13107 | 1893 |
| Attack | 7190 | 4232 |   | Attack | 103 | 7087 |   | Attack | 96 | 7 |   | Attack | 96 | 11326 |

## Permutation 2

| Siamese-NN | Normal | Attack |   | XGBoost | Normal | Attack |   | DNN | Normal | Attack |   | P2 | Normal | Attack |
|---|---|---|---|---|---|---|---|---|---|---|---|---|---|---|
| Normal | 14097 | 903 |   | Normal | 13414 | 683 |   | Normal | 13107 | 307 | = | Normal | 13107 | 1893 |
| Attack | 7190 | 4232 |   | Attack | 148 | 7042 |   | Attack | 96 | 52 |   | Attack | 96 | 11326 |

## Permutation 3

| DNN | Normal | Attack |   | Siamese-NN | Normal | Attack |   | XGBoost | Normal | Attack |   | P3 | Normal | Attack |
|---|---|---|---|---|---|---|---|---|---|---|---|---|---|---|
| Normal | 14378 | 622 |   | Normal | 13588 | 790 |   | Normal | 13107 | 481 | = | Normal | 13107 | 1893 |
| Attack | 103 | 11319 |   | Attack | 103 | 0 |   | Attack | 96 | 7 |   | Attack | 96 | 11326 |

## Permutation 4

| DNN | Normal | Attack |   | XGBoost | Normal | Attack |   | Siamese-NN | Normal | Attack |   | P4 | Normal | Attack |
|---|---|---|---|---|---|---|---|---|---|---|---|---|---|---|
| Normal | 14378 | 622 |   | Normal | 13755 | 623 |   | Normal | 13107 | 648 | = | Normal | 13107 | 1893 |
| Attack | 103 | 11319 |   | Attack | 96 | 7 |   | Attack | 96 | 0 |   | Attack | 96 | 11326 |

## Permutation 5

| XGBoost | Normal | Attack |   | Siamese-NN | Normal | Attack |   | DNN | Normal | Attack |   | P5 | Normal | Attack |
|---|---|---|---|---|---|---|---|---|---|---|---|---|---|---|
| Normal | 14129 | 871 |   | Normal | 13414 | 715 |   | Normal | 13107 | 307 | = | Normal | 13107 | 1893 |
| Attack | 481 | 10941 |   | Attack | 148 | 333 |   | Attack | 96 | 52 |   | Attack | 96 | 11326 |

## Permutation 6

| XGBoost | Normal | Attack |   | DNN | Normal | Attack |   | Siamese-NN | Normal | Attack |   | P6 | Normal | Attack |
|---|---|---|---|---|---|---|---|---|---|---|---|---|---|---|
| Normal | 14129 | 871 |   | Normal | 13755 | 374 |   | Normal | 13107 | 648 | = | Normal | 13107 | 1893 |
| Attack | 481 | 10941 |   | Attack | 96 | 385 |   | Attack | 96 | 0 |   | Attack | 96 | 11326 |

Figure 14: Permutations of Layer 1 classifiers for CIDDS-001 dataset